\pdfoutput=1
\documentclass[aps,preprint,floatfix,prl]{revtex4-1}
\usepackage{amsmath}
\usepackage{amsfonts}
\usepackage{amssymb}
\usepackage{graphicx}
\usepackage{booktabs}
\usepackage{color}
\usepackage{longtable}
\usepackage{array}
\usepackage{amssymb}
\usepackage[usenames,dvipsnames,svgnames,table]{xcolor}

\graphicspath{ {../Paper/images/}{../Paper/images/Growth_Rate_Cont_Plots/} }
\newcommand{\bx}{{\boldsymbol{\hat{x}}}}

\newcommand{\bu}{\mathbf{u}}
\newcommand{\grad}{\mathbf{\nabla}}

\newcommand{\hg}{h_g}
\newcommand{\Rey}{{R}}
\newcommand{\Ndg}{\tilde{N}_g}
\newcommand{\monami}{\textit{monami}}
\newcommand{\ubl}{U_\text{bl}}

\newcommand{\ReyNdg}{{\Rey\Ndg}}

\definecolor{tableShade}{gray}{0.8}
\title{Linear stability analysis for \textit{monami} in a submerged sea grass bed}
\begin{document}
\author{Ravi Singh}
\affiliation{Brown University, Providence RI 02912 USA}
% \author{L. Mahadevan}
% \affiliation{Harvard University, Cambridge MA 02138 USA}
\author{M. M. Bandi}
\affiliation{OIST Graduate University, Okinawa 904-0495, Japan}
\author{Amala Mahadevan}
\affiliation{Woods Hole Oceanographic Institution, Woods Hole MA 02543 USA}
\author{Shreyas Mandre}
\affiliation{Brown University, Providence RI 02912 USA}

\begin{abstract}
The onset of \monami ~-- the synchronous waving of sea grass beds driven by a steady flow -- is modeled as a linear instability of the flow. Unlike previous works, our model considers the drag exerted by the grass in establishing the steady flow profile, and in damping out perturbations to it. We find two distinct modes of instability, which we label Mode 1 and Mode 2. Mode 1 is closely related to Kelvin-Helmholtz instability modified by vegetation drag, whereas Mode 2 is unrelated to Kelvin-Helmholtz and arises from an interaction between the flow in the vegetated and unvegetated layers. The vegetation damping, according to our model, leads to a finite threshold flow for both these modes. Experimental observations for the onset and frequency of waving compare well with model predictions for the instability onset criteria and the imaginary part of the complex growth rate respectively, but experiments lie in a parameter regime where the two modes can not be distinguished. % The inclusion of vegetation drag differentiates our mechanism from the previous linear stability analyses of \monami.
\end{abstract}

\maketitle
\section{Introduction}
Sea grasses exhibit a rich set of dynamical behaviors due to their collective interaction with fluid flows.  
The hydrodynamic processes resulting from this behavior influence many environmental processes such as sedimentation, transport of dissolved oxygen, plant growth, and biomass production  \citep{Fonseca87,Grizzle96,Nepf99,Nepf2012}. 
One response of the submerged grass beds to steady currents is the formation of coherent large amplitude oscillations, known as \monami ~\citep{AckermanOkubo93}.  
In this article, we present a linear hydrodynamic instability underlying the onset of these coherent oscillations.

Current explanations of \monami~\citep{Ikeda96,Ghisal02,Raupach96} invoke the existence of a shear layer at the top of the grass bed (henceforth called grass top) due to vegetation drag. 
Its instability, through a mechanism similar to the Kelvin-Helmholtz (KH) instability, is thought to lead to coherent eddies over the grass bed.
The grass responds to these eddies by deforming, which leads to large amplitude synchronous oscillations of the grass blades.
% Basic application of this model predicts the observed frequency of \monami~and helps understand transport in the seagrass bed \citep{Nepf00,Ghisal02,Nepf04,Okamoto12}.
This picture can be used to derive a simple scaling dependence of the \monami~frequency on the flow speed and the shear layer thickness, and understand transport in the seagrass bed  \citep{Nepf00,Ghisal02,Nepf04,Okamoto12}.

\begin{figure}
\centerline{\includegraphics[scale=.99]{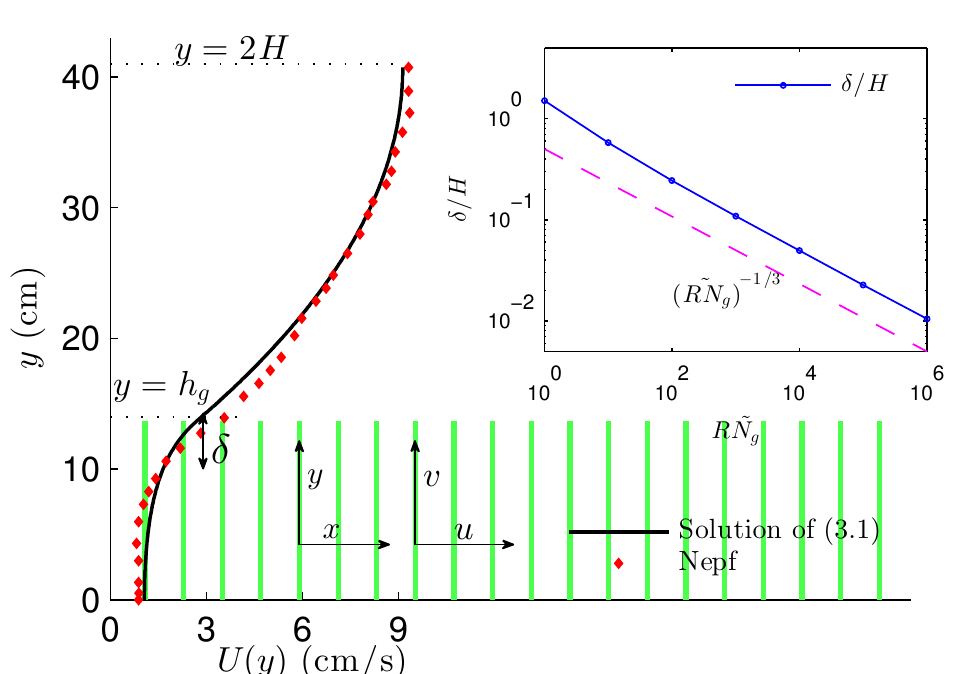} }
\caption{
Schematic setup and comparison of our steady flow profile with that from the experiments in ref. \cite{Nepf04} (Case I from Table 1) %  with 1250 plants/m$^2$, plant height = 13.7$\pm 0.2$ cm and blade width of 0.64 cm)
 and its approximation with $U_0=7.28$ cm/s and $\delta = 5.02$ cm in our model. The grass extends up to $y=\hg$ in the water column of depth $2H$. 
The steady velocity profile can be decomposed into a parabolic profile in the unvegetated region, a uniform profile deep within the vegetation, and a boundary layer of thickness $\delta$ near the grass top. 
The dependence of the boundary layer thickness (estimated as $|U/U_y|$ at $y=\hg$ from the numerical solution of \eqref{base_equ}) on the vegetation density parameter $\ReyNdg$ is shown in the inset.
}
\label{basicflow}
\end{figure}
However, several aspects of this explanation remain unsatisfactory. 
The instability is modeled using the inviscid Rayleigh equation and despite the role of drag in producing the shear layer, its role in damping the perturbations is ignored \citep{Raupach96}. 
Furthermore, the shear layer is assumed not to be influenced by the top and bottom boundary of the domain, although the experimentally measured thickness of the shear layer is in many cases comparable to the unvegetated water depth.
The velocity profile of the free shear layer is assumed \textit{ad hoc} to be piecewise linear \citep{Delangre06} or hyperbolic tangent \citep{Ghisal02,Raupach96}, with parameters fitted experimental observations.
The origin of these profiles, the values of the fitted parameters, and their effect on \monami~remains unexplained.  
% Assuming the profiles \textit{ad hoc} leaves the dependence of the shear layer thickness on the parameters of the problem unexplained.
And finally, no existing theory explains the threshold flow speed, observed in the lab \citep{Ghisal02} and the field \citep{Grizzle96}, below which \monami ~is not observed.

Here we present a mathematical model for the linear instability that accounts for these effects.
Although \monami ~is manifest in the motion of the grass, the drag exerted by the vegetation on the flow is central to the hypothesized instability. 
The instability and the resulting flow structures persist in lab experiments even when flexible grass mimics are replaced by rigid dowels \citep{Ghisal02,Nepf06}. 
Therefore, to develop the essential mathematical model, we assume the grass blades to be rigid, and oriented vertically (along $y$ direction) on average.
The vegetation is modeled as a continuum drag on the fluid acting perpendicular to the blade orientation and proportional to the grass density.
In the limit of dense vegetation, the steady profile established in the presence of the drag exhibits a localized region of enhanced shear gradient near the grass top and drives a flow instability.
% provides an opportunity to compare the instability mechanism with KH.
% The grass bed does not impose an effective no-slip boundary condition on the unvegetated flow; instead the flow penetrating the grass bed drives an instability.
While some features of instability are similar to KH, we also find significant differences.
This comparison is the focus of this article.

% We solve for the steady fully-developed flow profile established due to an imposed pressure gradient in the presence of this drag, and perform linear stability analysis of this profile.
% We compare the results of our calculations with data available from lab experiments and field observations.
% To better understand the flow instability mechanisms and to compare them with the current explanations of \monami, we investigate the limit of asymptotically dense vegetation. 
% 
% We find two modes of instability labeled Mode 1 and Mode 2.  
% We find that Mode 1 resembles many characteristics of KH instability but is modified by the presence of the vegetation drag. 
% We also find another mode of instability, Mode 2, unrelated to the KH instability mechanism, arising purely due to the interaction of an inviscid layer of fluid with submerged rigid vegetation.
% The presence of dissipation due to vegetation drag leads to a finite threshold flow condition for onset of the instability for both these modes.

\section{Mathematical model}
\begin{figure}
\begin{center}
\includegraphics[scale = 0.95]{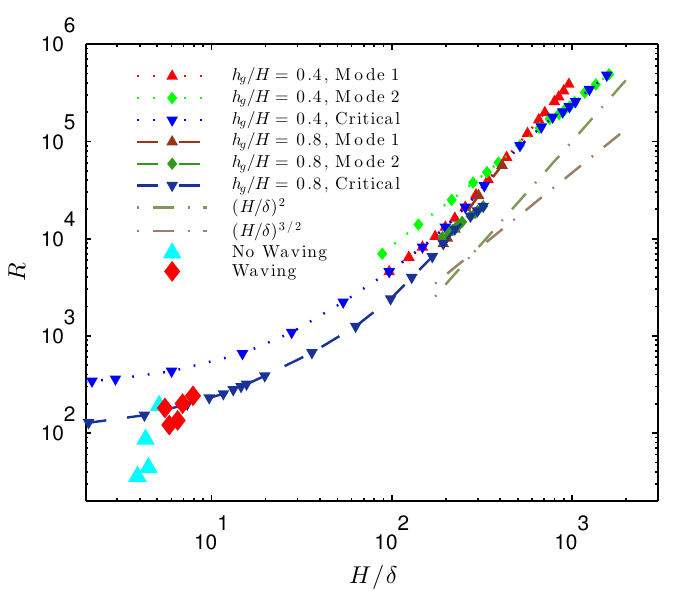}\includegraphics[scale = 0.95]{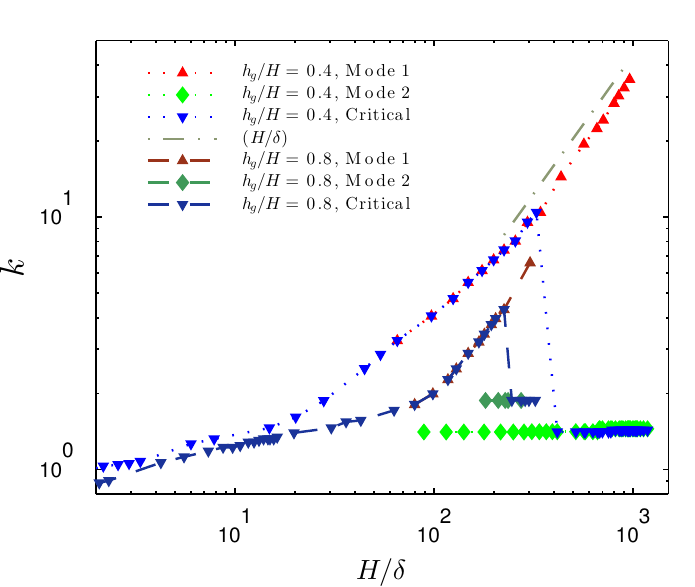}
% \includegraphics[width=12cm]{Critical_Re_vs_delta_noshear} \\
% \vspace{-6mm} \hspace{-5mm}
% \includegraphics[width=12cm]{K_vs_shear_width_noshear}
\end{center}
\caption{
Critical Reynolds number, threshold Reynolds number for Mode 1 and Mode 2 (left) and the corresponding marginally stable wave number (right) for different submergence ratio as a function of vegetation density parametrized by the boundary layer thickness. 
Parameters from experiments reported by \cite{Ghisal02} to exhibit or suppress synchronous waving are also included in the left panel. 
In order to estimate the $\Rey$ for these experiments, a representative value of $\mu=0.1$ Pa~s was assumed.
}
\label{Re_vs_delta}
\end{figure}
%We show that the drag of the grass results in an instability mode different from the KH instability.
The vegetation is assumed sufficiently dense so that the drag exerted by it may be modeled by a continuous body force $\mathbf{f}$ entering the equation governing the flow as.
\begin{equation}
\grad\cdot\bu = 0,\hspace{3mm} \rho \left(\bu_{t}+\bu.\grad\bu \right) = -\grad p+\mu\grad^{2}\bu +\mathbf{f}, %+\rho\mathbf{g}
\label{navier-stokes}
\end{equation}
where $\rho$, $\bu=(u,v)$, and $p$ are the fluid density, velocity, and the dynamic pressure respectively, and $\mu$ is the (dynamic) eddy viscosity. % and $\mathbf{g}$ the acceleration due to gravity. 
The Reynolds number of the flow based on the scale of the grass blade is $O(10^2-10^3)$; therefore, neglecting skin friction, we model the form drag on the vegetation as $\mathbf{f}=-N_g C_N \rho u |u| d\bx$ \citep{Nepf99,Nepf00,Nepf04}, where  $N_g$ is the number of grass blades per unit horizontal area, $C_{N}$ the drag coefficient for the flow normal to the grass, and $d$ the average blade width projected perpendicular to the flow. 
In the interest of simplicity, we model the turbulence using an eddy viscosity; see \S\ref{subsec:turbulencemodel} for a detailed discussion. 
In the field, $C_N$, $N_g$ and $\mu$ vary with position, but we do not expect these variations to be central to the instability mechanism, and therefore take them to be constants. 
% Based on previous experiments \cite{Vivoni98,Nepf00}, we take $C_N = 1$.
% Note that only the product $C_N N_g d$ appears together in our model. 

\section{Linear stability analysis}
We first calculate the fully developed steady solution $\bu = U(y)\boldsymbol{\hat{x}}$ of ~\eqref{navier-stokes} driven by constant pressure gradient $dP/dx$ in a water column of depth $2H$ and vegetated depth $\hg$, and use it to non-dimensionalize the mathematical model. The momentum balance \eqref{navier-stokes} for $U(y)$ simplifies to
\begin{equation}
 -\frac{dP}{dx}+\mu U''(y) -S(y) \rho C_N d N_gU |U|=0,
\label{base_equ}
\end{equation}
where $S(y)=1$ for $0<y<\hg$ and $S(y)=0$ for $\hg< y< 2H$. 
Eq. \eqref{base_equ} is solved subject to no shear at the boundaries, i.e., $U'(0) = U'(2H) = 0$.
The lower boundary conditions is appropriate for dense vegetation because the shear stress exerted by the bottom surface is expected to be negligible compared to the vegetation drag~\citep{Nepf00}, whereas the upper boundary condition models the free interface. 
A comparison of the steady flow profile from the solution of ~\eqref{base_equ} with experimental measurements by \cite{Nepf04} is shown in Fig.~\ref{basicflow}.
The profile $U(y)$ has three distinct regions.
Within the vegetation, it is approximately uniform with $ U(y) \approx U_g = \sqrt{\frac{-dP/dx}{\rho C_N dN_g}}$, arising from a balance of the drag with the pressure gradient. 
Outside the vegetation, the velocity has a simple parabolic profile. % due to the balance between viscous forces and the pressure gradient. 
At the grass top, continuity of shear stresses results in a boundary layer of thickness $\delta$. 
Denoting $\ubl$ to be the velocity scale in the boundary layer, and $U_0 = {(-dP/dx)~H^2}/{\mu}$ the velocity scale in the unvegetated region, the balance between viscous forces and vegetation drag implies $\mu \ubl/\delta^2 \sim \rho C_N d N_g \ubl^2$, and the continuity of shear stress across the grass top implies $\ubl/\delta \sim U_0/H$.
Solving for $\delta$ and $\ubl$ yields $\delta/H \sim \ubl/U_0 \sim (\ReyNdg)^{-1/3}$, where $\Ndg = \left(C_N d H N_g\right)$ is the vegetation frontal area per bed area, and $\Rey=\rho U_0 H/\mu$ is the Reynolds number of the flow. 
A numerical estimate of $\delta$ (estimated as $U/U_y$ at $y=\hg$) is compared with this prediction in Fig. \ref{basicflow} (inset).
We identify the boundary layer to be analogous to the shear layer \citep{Ghisal02,Nepf04} in the previous explanation of \monami.
This dependence of $\delta$ on $N_g$ gives us a way to systematically investigate the effect of the shear layer thickness on the instability mechanism.
Fig. \ref{basicflow} also shows that the asymptotic regime of a thin boundary layer is expected to hold for $\ReyNdg \gtrsim 100$. 
In this notation, $U_g/U_0 = (\Rey \Ndg)^{-1/2}$ (used later in deriving \eqref{eqn:mode2asymp}). 
\begin{figure}
\centerline{\includegraphics[]{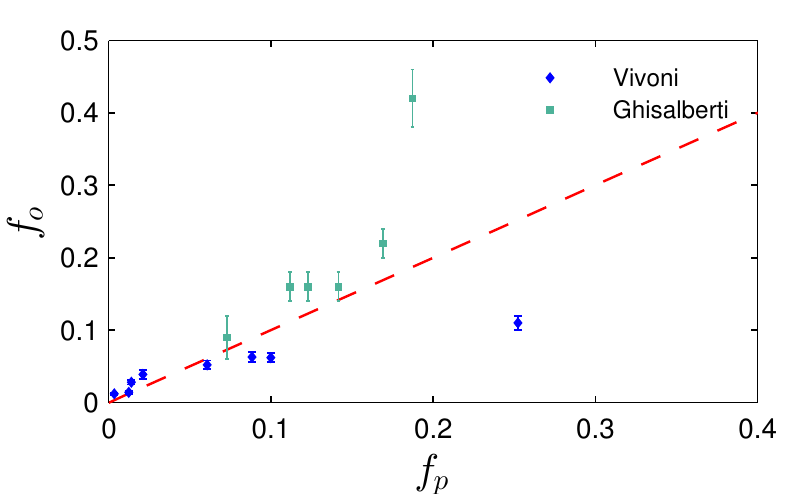}}
\caption{Comparison of experimental observations of the experimentally measured dominant frequency $f_o$ (in Hz) with the predictions $f_p=\text{Im}(\sigma)$ from the solution of ~\eqref{Orr-somerfield}. 
The experimental data in the inset is obtained from \cite{Ghisal02} and \cite{Vivoni98}. 
In order to estimate the $\Rey$ for these experiments, a representative value of $\mu=0.1$ Pa~s was assumed.
}
\label{frequency_comparison}
\end{figure}

Next we substitute $\bu = (U+\tilde{u}, \tilde{v})$, $p=P+\tilde{p}$ in ~\eqref{navier-stokes} and expand to linear order to investigate the evolution of small perturbations $(\tilde{u}, \tilde{v})$, which obey
\begin{equation}
\begin{split}
\rho(u_t+U u_x+vU_y) &= -p_x+ {\mu}\nabla^2u-2S\rho C_{N}dN_{g}Uu, \\
\rho(v_t+ Uv_x) &= -p_y+ {\mu}\nabla^2v, \hspace{0.3cm} \nabla\cdot\bu=0,
\end{split} \nonumber
\end{equation}
where the tilde are dropped.
These equations are non-dimensionalized using half channel height $H$, velocity $U_0$, and time $H/U_0$, leading to three non-dimensional parameters, \textit{viz.} $\Rey$, $\Ndg$, and the vegetation submergence ratio $\hg/H$. 
We also use $\delta/H$ in lieu of $\Ndg$ to parametrize the vegetation density and help elucidate the instability mechanism. 
Using a stream function $\psi$ with $u = \psi_{y}, v= -\psi_x$ to satisfy mass balance, we seek a solution of 
the form $\left(u,v,\psi \right)= \left(\hat u(y), \hat v(y), \phi(y) \right)e^{ikx+\sigma t}$ to obtain a modified Orr-Sommerfeld equation \citep{Drazin81,Chen97,Chu91} 
\begin{equation}
\begin{split}
\Rey^{-1}\left(D^2 -k^{2} \right)^2\phi &= \left[ \left({\sigma}+ikU\right) \left(D^2-k^2\right) -ikU_{yy}\right]\phi + \Ndg D\left(2 S U D \phi\right),
\label{Orr-somerfield}
\end{split}
\end{equation}
where $D=d/dy$, and subject to the boundary conditions $\phi = D^2\phi = 0$ at $y=0$ and $y=2$. 
The growth rate $\sigma$ for a given wave number $k$ appears as an eigenvalue that allows a non-trivial solution $\phi$ of  \eqref{Orr-somerfield}.
We solve \eqref{Orr-somerfield} numerically for $\sigma$ and $\phi$.

\section{Results - unstable modes and critical parameters}
A threshold in $\Rey$, above which the flow is unstable (Re$(\sigma)>0$) for at least one $k$, emerges from the solution of ~\eqref{Orr-somerfield}. 
The dependence of this threshold $\Rey$, and the corresponding marginally stable wavenumber $k$, on $\delta/H$ and $\hg/H$ is shown in Fig.~\ref{Re_vs_delta}, and is found to compare well with experimental observations of \cite{Ghisal02}.
The threshold $\Rey$ increases with the vegetation density, indicating a competition between the destabilizing shear in the flow, and the stabilizing effect of damping due to vegetation drag.
% A similar conclusion was presented for an analogous problem (flow around an emergent (i.e., $\hg>2H$) sea grass patch), but by assuming $U(y)$ to be a tanh-profile, and neglecting the viscous term \citep{White07}.
%Previous calculations for terrestrial grass either exclude the vegetation drag in their models \cite{Raupach96}, or assume the mean velocity profile \textit{ad hoc} \citep{Raupach96,Delangre06}.
%A threshold flow condition is not reported previously either for terrestrial or submerged marine meadows.
The frequency (Im$(\sigma)$) of the fastest growing mode also agrees well with observed behaviour -- frequency of \monami, maxima in the velocity spectra, and frequency of vortex passage in lab scale experiments \citep{Ghisal02} -- for cases where the vegetation was sufficiently dense to be modeled by a continuum drag field as shown in Fig.~\ref{frequency_comparison}.  
The experimentally observed \monami ~wavelengths are not available for comparison.

To better understand the instability mechanism, we consider the dependence of the fastest growing wavenumber on $\delta$.
The fastest growing wavenumber first increases proportional to $H/\delta$, but at a critical $\delta$ discontinuously jumps and remains $O(1)$ (see Fig.~\ref{Re_vs_delta}). 
To aid in explaining this behavior, we show heat maps of Re$(\sigma)$ as a function of $\Rey$ and $k$, for different $\hg/H$ and $\Ndg$ in Fig.~\ref{K_Re_sigma_set3}. 
The smallest $\Rey$ on the neutral curve (Re$(\sigma)=0$) sets the threshold. 
We observe that as $\Ndg$ increases, the unstable region splits into two; we refer to the region with the higher $k$ as ``Mode 1'', and the one with the lower $k$ as ``Mode 2''. 
For $\hg/H\lesssim 0.9$, the unstable region for Mode 1 recedes to higher $\Rey$, and for $\hg/H \gtrsim 0.9$, the region shrinks to zero size.
In either case, due to such behaviour the most unstable mode transitions discontinuously from Mode 1 to Mode 2.
All experimental data we have found corresponds to a vegetation density for which the unstable region in the $\Rey-k$ space has not split into two. 

\section{Discussion -- comparison of unstable modes with Kelvin Helmholtz}
The distinct asymptotic behavior of the two modes as $\Ndg \gg 1$ distinguish them from each other and facilitate comparison with KH instability mechanism. 
\subsection{Mode 1}
We numerically observe that the threshold Reynolds number for Mode 1 instability scales as  $\Rey \sim (H/\delta)^2$ (or $\Rey \propto {\Ndg}^{2}$). 
Our calculations also show that Mode 1 asymptotically localizes to the boundary layer near the grass top, and exhibits highest growth for a perturbation of  $k \sim H/\delta$ (see Fig.~\ref{Re_vs_delta}). 
The behavior of this critical Reynolds number can be understood by considering the limit $\Rey \gg 1$ and $\Ndg \gg 1$.
Estimating the size of various terms of ~\eqref{Orr-somerfield} within the boundary layer in this limit help us understand the behavior of the critical $\Rey$. 
Using $D\sim H/\delta$, $\sigma \sim O(1)$, and $U=\ubl \sim \delta/H$ in the boundary layer; the magnitude of the advection term is $ (H/\delta)^2$  (or $\Rey^{2/3} \Ndg^{2/3}$), and the viscous and vegetation drag terms are $(1/\Rey) (\delta/H)^{-4}$ (or $(\Rey^{1/3} \Ndg)^{4/3})$. 
The advection term, viscous term and vegetation drag terms balance when $\Rey \sim (H/\delta)^2$ (or $\Rey \sim {\Ndg}^{2}$).

To further understand the mechanism of Mode 1, we rescale \eqref{Orr-somerfield} near the grass-top using the the boundary layer scalings $\eta = y/(\delta/H)$, 
$U(y) = (\delta/H)\bar{U}(\eta)$ and $k = (H/\delta) \bar{k}$.
With these scalings ~\eqref{Orr-somerfield} simplifies to
\begin{equation}
\begin{split}
\left(\bar{D}^2 -\bar{k}^{2} \right)^2\phi &= (\Rey/\Ndg^2)^{1/3} \left[ \left({\sigma}+i\bar{k}\bar{U}\right) \left(\bar{D}^2-\bar{k}^2\right) -i\bar{k}\bar{U}_{\eta\eta}\right]\phi + \bar{D}\left(2S \bar{U} \bar{D} \phi\right),
\label{eqn:mode1asymp}
\end{split}
\end{equation}
in a region of thickness O($\delta$) near $y=\hg$, where $\bar{D} = d/d\eta$. 
Since $(\Rey/\Ndg^2)$ is the only remaining parameter in \eqref{eqn:mode1asymp}, the mode shape and solution are expected to converge in the limit $\Rey \gg 1$, $\Ndg \gg 1$, but $\Rey/\Ndg^2$ fixed.
Our numerical findings confirm this expectation; the critical $\Rey$ scales as $(H/\delta)^2$ as shown in Fig.~\ref{Re_vs_delta} and the mode shapes are self-similar with length scale $\delta$, as shown in Fig \ref{Asymptotic_mode}. 

Mode 1 shares many characteristics with the KH instability (see Table \ref{tab:comparison}). 
The fastest growing wavenumber at the critical $\Rey$ scales as $k \propto (H/\delta)$, similar to KH instability. 
The extent of the unstable mode is also localized to the boundary layer region.
The porous nature of the vegetation implies that a weak flow of magnitude $\ubl = U_0 \delta/H$ penetrates a thin boundary layer region $\delta$, and therefore the shear gradient $U_{yy} \sim U_0/\delta H$ is largest in this region. 
The strong shear gradient $U_{yy}$ in the boundary layer plays a central role in destabilizing the flow and localizing the instability to that region. 
\begin{figure*}
 \includegraphics[width=\textwidth]{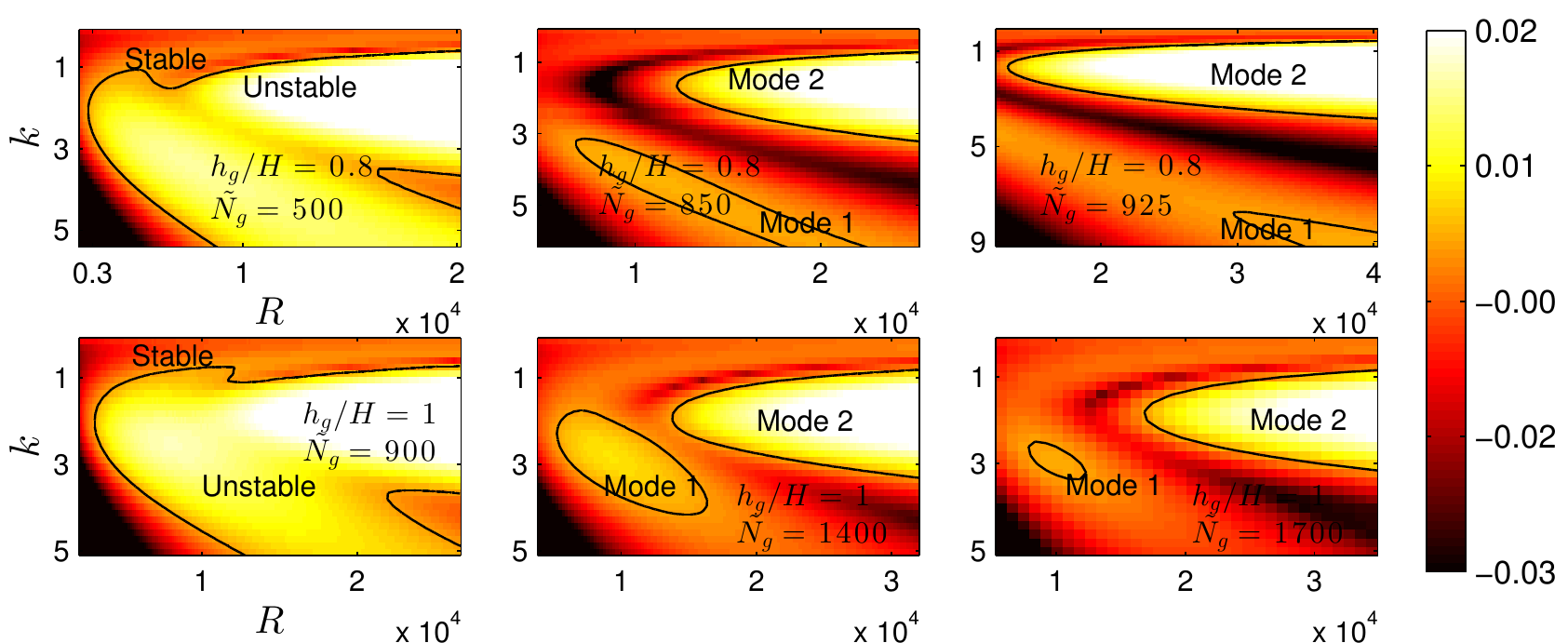}
%\end{figure}
%\begin{figure*}
%\begin{tabular}{cccc}
%{\includegraphics[height = 4cm, width = 5.8cm]{Set4_dens28_imgsc}} &
%{\includegraphics[scale = 0.67]{Set4_dens30_imgsc}} &
%{\includegraphics[height = 4cm,width = 5.8cm]{Set4_dens32_imgsc}} &
%{\includegraphics[height = 4.15cm,width=6.5cm]{Set4_dens34_imgsc}} \\
%{\includegraphics[height = 4cm, width = 5.8cm]{Set5_dens38_imgsc}} &
%{\includegraphics[scale = 0.67]{Set5_dens40_imgsc}} &
%{\includegraphics[height = 4cm, width = 5.8cm]{Set5_dens42_imgsc}} &
%{\includegraphics[height = 4.15cm,width=6.5cm]{Set5_dens46_imgsc}} \\

%{\includegraphics[scale = 0.67]{Set6_dens32_imgsc}} &
%{\includegraphics[scale = 0.67]{Set6_dens34_imgsc}} &
%{\includegraphics[scale = 0.67]{Set6_dens36_imgsc}} &
%{\includegraphics[scale = 0.67]{Set6_dens38_imgsc}}
%\end{tabular}
\caption{
$\text{Re}(\sigma)$ and the neutral curve ($\text{Re}(\sigma)$=0) as a function of wavenumber and $\Rey$ for parameters shown in the corresponding panel.  
As $\Ndg$ increases, the unstable region splits into two labeled as ``Mode 1'' and ``Mode 2''. 
For $\Ndg$ below (above) a critical value, Mode 1 (Mode 2) sets the threshold $\Rey$.
}
\label{K_Re_sigma_set3}
\end{figure*}

Our detailed description of Mode 1, given by \eqref{eqn:mode1asymp} also highlights key differences with formulations of KH. 
KH is usually described using the inviscid Rayleigh's equation, 
\begin{align}
\left(\sigma+ikU\right) \left(D^2-k^2\right)\phi =  ikU_{yy}\phi, 
\label{eqn:Rayleigh}
\end{align}
and is therefore not parametrized by the Reynolds number.
Describing the instability using the Orr-Sommerfeld equation introduces the Reynolds number as a parameter, but shear flows with tanh-profiles are unstable for all values of the parameter \citep{Drazin81}.
Therefore, based on the inviscid formulations of KH instability, the origin of the threshold flow conditions observed in experiments and the field is unclear.

In our model, (turbulent eddy) viscosity sets the scale of the boundary layer, and therefore for Mode 1.
However, the boundary layer is established only in the vegetated region; the velocity profile does not saturate on the scale of $\delta$ in the unvegetated region.
The threshold flow condition arises from a competition between the destabilizing role of fluid inertia, which is very similar to the one played in KH, and the vegetation drag.
The vegetation drag may not be neglected within this boundary layer, and therefore plays a central role in the Mode 1 instability mechanism.

%We expect identical asymptotic behavior for a fixed relative magnitude of the terms on the r.h.s. to those on the l.h.s. of \eqref{eqn:mode1asymp}, which is $(\Rey \delta/H)^{3/2}$ (or $\Rey/\Ndg^{1/2}$).
%Therefore the threshold obtained for Mode 1 is $\Rey \propto H/\delta$ (or $\Rey \propto \Ndg^{1/2}$) explaining the numerically observed asymptote (see Fig.~\ref{Re_vs_delta}). 
%This analysis also concludes that the mode structure is self-similar over the length scale $\delta$ for fixed $\Rey/\Ndg^{2}$; the verification of this idea is shown in Fig.~\ref{Asymptotic_mode} (inset).

\subsection{Mode 2}
\begin{figure}
\centerline{\includegraphics[width=9cm]{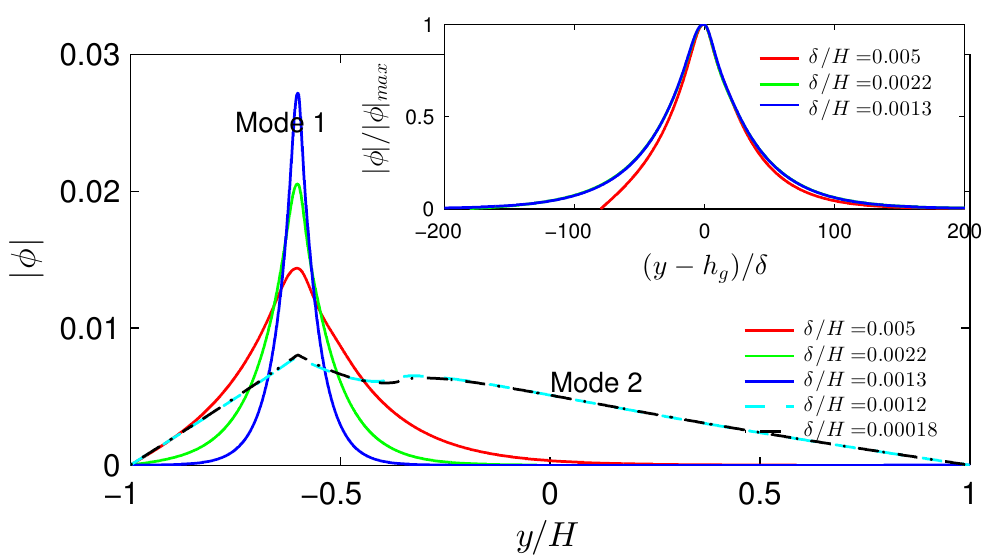}}
\caption{
Plot of the neutral Mode 1 (solid) and Mode 2 (dashed) shape $|\phi|$ in the limit of small $\delta/H$ for $\hg/H=0.2$. 
% Mode 1 is shown in solid and Mode 2 is shown in dashed. The parameters for Mode 2 shapes are chosen such that $\Rey \gg 1$, $\Ndg \gg 1$ (specified in terms of $\delta/H$) but $\Rey/\Ndg = O(1)$. 
The approach of mode shapes to each other for these small values of $\delta/H$ indicates that the dense vegetation asymptote is reached. 
Mode 1 shapes appear self-similar in shape as $\delta\to 0$.
Inset shows rescaled $|\phi|$ for Mode 1 as a function of $(y-\hg)/\delta$ approach a universal shape, indicating that an asymptotic limit has been reached. 
% The limit is not yet reached for the case $\delta/H = 0.005$ due to the influence of bottom boundary; the vegetation height in this case is comparable to the boundary layer thickness.
}
\label{Asymptotic_mode}
\end{figure}

The threshold condition for Mode 2 is numerically observed to be $\Rey \propto ({\delta}/{H})^{-3/2}$ (or $\Rey \propto \Ndg$) for $k\sim O(1)$, shown in Fig.~\ref{Re_vs_delta}, which can be understood by assuming $\Rey \gg 1$ but fixed $\Rey/\Ndg \sim O(1)$.
% Note that for large $\Ndg$, the non-dimensional steady flow velocity inside the vegetation is $U_g \sim (\Rey \Ndg)^{-1/2}$. 
In this limit, the non-dimensional flow in the grass bed is $U_g/U_0 \sim (\Rey \Ndg)^{-1/2} \ll 1$, and therefore $ikU \ll \sigma$ may be neglected in comparison to $\sigma$. 
Furthermore, $U_{yy}$ decays to zero within the grass outside the boundary layer . 
Outside the grass, the turbulent viscous stress term is negligible compared to the inertial term because $\Rey \gg 1$. 
Thus, \eqref{Orr-somerfield} simplifies to 
\begin{subequations}
\begin{align}
% \begin{split}
\sigma\left(D^2-k^2\right)\phi = -2{(\Ndg/\Rey)^{1/2}}D^2\phi,  \quad &\text{ for } y<\hg  \label{eqn:mode2asympa} \\
\left(\sigma+ikU\right) \left(D^2-k^2\right)\phi =  ikU_{yy}\phi, \quad &\text{ for } y>\hg. \label{eqn:mode2asympb}
% \end{split}
\end{align}
\label{eqn:mode2asymp}
\end{subequations}
The only remaining parameter in \eqref{eqn:mode2asymp} is $\Rey/\Ndg$. 
For fixed $\Rey/\Ndg$, the mode shape converges in the aforementioned limit, in agreement with our numerical results shown in Fig.~\ref{Asymptotic_mode}.
This convergence indicates that we have identified the correct asymptotic limit to investigate Mode 2.
The parameter $\Rey/\Ndg$ therefore sets the threshold, justifying the numerically observed asymptotic behavior $\Rey \propto \Ndg$ (or $\Rey \sim ({\delta}/{H})^{-3/2}$; see Fig.~\ref{Re_vs_delta} for comparison with numerical results).

The structure of this mode in the aforementioned limit is such that $\phi$ is continuous at $y=h_g$, but $D\phi$ undergoes a rapid transition there, on the scale of boundary layer thickness $\delta$.
The eigenvalues and the mode shape are otherwise independent of $\delta$.
Therefore we conclude that the boundary layer only plays a secondary role of regularizing the discontinuity in tangential velocity arising at $y=\hg$ in this instability mechanism.
The enhanced shear in the boundary layer plays no role for this mode of instability.
% We interpret Mode 2 as the instability of an inviscid fluid, with the vegetation modeled by a continuum drag field, and for which the boundary layer near the grass top plays no role. 
% On the other hand, Mode 1 asymptotically localizes to the boundary layer near the grass tip, and exhibits a different asymptotic behavior with $k \sim O(H/\delta)$, and $\Rey \sim (H/\delta)$ (or $\Rey \propto {\Ndg}^{1/2}$) at the threshold. 

Mode 2 has characteristics distinct from KH. Outside the grass, the unstable mode shape is governed by the inviscid Rayleigh's equation \eqref{eqn:Rayleigh}.
An inflection point in $U(y)$ is a necessary condition for instability arising from \eqref{eqn:Rayleigh} according to Rayleigh's criteria \citep{Rayleigh1879}. 
However, for our $U(y)$ profiles, $U_{yy}(y) = -1$ above the grass and therefore does not change sign for $y>\hg$. 
Instead, the dynamics are coupled with the flow in the grass bed described by \eqref{eqn:mode2asympa} in $y< \hg$.
The absence of $U_{yy}$ in \eqref{eqn:mode2asympa} indicates that $U_{yy}$ is approximated to be zero in $y<\hg$, and therefore the positive values of $U_{yy}$ that occurs in the boundary layer do not affect this mode of instability to leading order.
Furthermore, the presence of the critical parameter $\Rey/\Ndg$ in \eqref{eqn:mode2asympa} indicates the presence of alternative destabilizing dynamics, involving the interaction of flow in the unvegetated region governed by \eqref{eqn:Rayleigh} with the flow in the vegetated region incorporating the drag.
Therefore, we conclude that Mode 2 is distinct from the KH instability, and owes its existence to vegetation drag.

% \subsection{Comparison with KH}
% Table \ref{tab:comparison} compares the two modes to each other, and to the KH instability. 
% Because the eigenfunction of Mode 1 is localized over a length scale $\delta$, it may be interpreted as the instability of the flow in the boundary layer, whereas Mode 2 may be  % understood as the instability on the scale of the water column. 
% Mode 1 appears to be superficially similar to the Kelvin Helmholtz mechanism, whereas Mode 2 arises purely from the interaction between the unvegetated water column and the flow through the vegetation. 
% The appearance of the vegetation drag parameter in the dominant balances represented by ~\eqref{eqn:mode1asymp} and ~\eqref{eqn:mode2asymp}, and the resulting threshold criteria % demonstrates its role in setting the threshold, and in distinguishing them from the KH instability.
% Vegetation drag plays a dominant role in the mechanism for both the modes, which distinguishes our analysis from the traditional KH instability. 

\begin{table*}
% \footnotesize
\rowcolors{3}{tableShade}{white}  %% start alternating shades from 3rd row
\renewcommand{\arraystretch}{1.2}
 \begin{tabular}{l|c|c|c}
			& KH 				& Mode 1 		& Mode 2 \\ \hline
 Base velocity profile 	& $U(y) = U_0 \tanh(y/\delta)$			& \multicolumn{2}{c}{Equation \eqref{base_equ}} \\
 Domain 		& $-\infty < y < \infty$			& \multicolumn{2}{c}{$-1<y<1$} \\
 Inflection point	& exists at $y=0$				& \multicolumn{2}{c}{$U''(y)$ discontinuous at $y=\hg$} \\
 Shear layer thickness	& $\delta$					& \multicolumn{2}{c}{$\delta \sim  H\left(\Rey \Ndg \right)^{-1/3}$} \\
 Linearized dynamics	& Equation \eqref{eqn:Rayleigh}		& \multicolumn{2}{c}{Equation \eqref{Orr-somerfield}} \\
 Dense grass limit &  no grass included & Equation \eqref{eqn:mode1asymp} & Equation \eqref{eqn:mode2asymp}  \\
 Critical parameters	& none						& $\Rey \propto \Ndg^{2}$ 	& $\Rey \propto \Ndg$ \\
 Most unstable $k$ as $\delta \to 0$	& $\propto H/\delta$		& $\propto H/\delta$	& $O(1)$ \\
 Mode localized?	& yes, near $y=0$				& yes, near $y=\hg$			& no, spans water column
 \end{tabular}
 \caption{Comparison between KH instability and the two unstable modes resulting from solution of \ref{Orr-somerfield}.}
 \label{tab:comparison}
\end{table*}
\subsection{Role of turbulence model}\label{subsec:turbulencemodel}
We have modeled turbulence using a constant eddy viscosity. 
The simplicity of this model allows us to make progress and capture the essential features of the instability. 
% without accounting for the rich and detailed characteristics of the turbulence. 
However, this simplicity in some cases only provides a qualitatively accurate description of the flow. 
In this subsection, we present an account of the advantages and shortcomings of assuming a constant eddy viscosity to model the turbulence.

As a consequence of the constant eddy viscosity, the boundary layer thickness $\delta$ scales as $H(\ReyNdg)^{-1/3}$.
Experimental observations show that the boundary layer thickness scales instead as $\Ndg^{-1}$ \citep{Nepf07}.
Whereas the precise boundary layer thickness is governed by the details of the turbulence model, the existence of this boundary layer for dense vegetation is independent of the turbulence model. 
We have captured one possible realization of this feature using constant eddy viscosity.
Experiments have also shown that a model based on mixing length $l$ better approximates the turbulent characteristics of the flow with $l \sim \delta$; \textit{i.e.}, the boundary layer itself establishes eddies to transport momentum. 
The eddy viscosity corresponding to this model is $\mu \sim \rho U \delta$, and the leading order balance between turbulent momentum transport and vegetation drag is $\mu U/\delta^2 \sim \rho C_N d N_g U^2$.
Substituting $\mu$ yields $\delta/H \sim \Ndg^{-1}$, in agreement with the experimental observations.

% Replacing the constant eddy viscosity by the scale for it from the mixing length hypothesis also recovers the experimentally observed scaling for boundary layer thickness.
Within our framework, the mixing length model implies a scale for the eddy viscosity $\mu \sim \rho \ubl \delta$ at grass top, which corresponds to an effective $\Rey \sim U_0H/\ubl \delta$.
Furthermore, matching the slope of the velocity profile from the boundary layer to the unvegetated flow implies $\ubl/U_0 \sim \delta/H$, and therefore $\Rey \sim (H/\delta)^2$.
Substituting this relation in $\delta/H \sim (\Rey \Ndg)^{-1/3}$ and solving for $\delta$ yields $\delta/H \sim \Ndg^{-1}$.
This simple scaling analysis shows that the boundary layer thickness depends on the turbulence model, and indicates that turbulence models based on mixing lengths will yield more realistic scalings for boundary layer thickness.
At the same time, the qualitative features of the instability are represented by our analysis.

The Mode 1 instability is driven by the intense shear on the scale of the boundary layer.
The driving mechanism for this instability is similar to that of KH, and relies only on the presence of this shear as presented in the $\bar{U}_{\eta\eta}$ term in equation \eqref{eqn:mode1asymp}. 
Therefore, we expect Mode 1 instability to be exhibited independent of the turbulence model. 
We further expect the fastest growing wavenumber to be proportional to $1/\delta$, and the mode to be localized to the boundary layer because these results have a basis in dimensional analysis.
The threshold parameters for Mode 1, however, may depend on the precise turbulence model used.
 
For the Mode 2 instability, the turbulent momentum transport is found to be irrelevant to leading order.
In the asymptotic limit of dense grass, \eqref{eqn:mode2asymp} shows that the instability is driven by the interaction of the unvegetated flow with the vegetation drag.
The influence of the turbulence model is limited to the regularization of the sharp transition in tangential velocity across the grass top.
Therefore we expect Mode 2 and its features to be preserved even if a different turbulence model is used.

\subsection{Comparison with previous models}

% Chen and Jirka (1997), White and Nepf (2007). What are their results and how do they relate to ours? These analyses have investigated instability of hyperbolic tangent profiles using modified versions of Orr Sommerfeld or Rayleigh's equations in the context of shallow water flow instabilities. They have represented the additional drag arising due to vegetation or bottom friction using a parametrization analogous to ours. 
A modified version of the Orr-Sommerfeld equation was analyzed previously \cite{Chu91,Chen97,White07} in the context of instabilities in depth-averaged shallow water flows, where bottom friction replaces or augments vegetation drag.
They assumed the steady profile to be a hyperbolic tangent, the drag to be isotropic, and the flow domain to be infinite in $y$.
One study \cite{White07} also neglected the eddy viscosity in their stability analysis.
While a detailed investigation needed to compare the consequence of the different assumptions is outside the scope of this paper, we discuss similarities and differences between their results and ours.
These investigations only found one unstable mode.
It is most likely so because the calculations were restricted to a parameter regime where the two modes have not yet been separated from each other, as is the case shown in Fig. \ref{K_Re_sigma_set3} for the lowest $\Ndg$.
These investigations also found that increasing the drag could further destabilize the flow, which is consistent with our interpretation of the Mode 2 instability mechanism.

The analogous oscillation of terrestrial canopies in wind, known as \textit{honami} \citep{Inoue56,Raupach96}, is different because the atmospheric boundary layer is much larger than the vegetation height.
% A crucial difference between the atmospheric and aquatic flow is that the atmospheric flows are essentially unbounded vertically \cite{Vivoni98,Nepf00}. 
% Another difference is that terrestrial vegetation is much more rigid, whereas aquatic vegetation is buoyant \cite{Vivoni98,Ghisal02}. 
In the framework of our model, the limit of $\hg/H \ll 1$ while $\delta/\hg$ = constant can be used to represent the hydrodynamic instability for the terrestrial case.
We find that in this case, the transition from Mode 1 to Mode 2 happens at such a large vegetation density, that Mode 2 is irrelevant. 
Hence, only the KH-like characteristics are observed in the terrestrial case. 

% We now test the assumption of an undeformable grass bed due to the dominant restoring force of buoyancy, using the criteria that the buoyancy time scale be much shorter than the hydrodynamic time scale $H/U_0$.
% For a common seagrass, \textit{Zostera Marina}, the relative density difference $\Delta \rho /\rho \approx 0.25$, the volume fraction $V_f \approx 0.1$ and $H=1$ m \citep{Fonseca98}, yielding the buoyancy time scale $\sqrt{\rho H/V_f \Delta \rho g} \approx 2$ s.
% The hydrodynamic time scale assuming $U_0 \approx 0.1$ m/s is 10 s, and therefore longer than the hydrodynamic time scale.
% We have neither accounted for the pre-factors appearing in the scaling argument, or considered cases when the time-scale separation is not so evident.
% Accounting for these factors  can lead to further interesting behavior \citep{Delangre06}.

While predictions of the threshold $\Rey$ for the onset of the instability and the frequency of oscillations are comparable to experimental observations, the deviation of our model predictions from the observed may be attributed to the various simplifications in our model. 
In real meadows, the grass is flexible, and the drag coefficients are known to vary from bottom to tip of the grass blades due to variation in vegetation characteristics \citep{Vivoni98,Nepf00}. 
The turbulence model for the flow through the meadow can also be improved from one with constant eddy viscosity \citep{Ghisal02, Nepf04}. 
Although these model improvements might lead to a better agreement between the observed and the predicted quantities, the insight furnished by \eqref{eqn:mode1asymp} and \eqref{eqn:mode2asymp}, and therefore our main conclusions, remain useful.

\section{Conclusion}
% We now test the assumption of an undeformable grass bed due to the dominant restoring force of buoyancy, using the criteria that the buoyancy time scale be much shorter than the hydrodynamic time scale $H/U_0$.
% For a common seagrass, \textit{Zostera Marina}, the relative density difference $\Delta \rho /\rho \approx 0.25$, the volume fraction $V_f \approx 0.1$ and $H=1$ m \cite{Fonseca98}, yielding the buoyancy time scale $\sqrt{\rho H/V_f \Delta \rho g} \approx 2$ s.
% The hydrodynamic time scale assuming $U_0 \approx 0.1$ m/s is 10 s, and therefore longer than the hydrodynamic time scale.
% We have neither accounted for the pre-factors appearing in the scaling argument, or considered cases when the time-scale separation is not so evident. Accounting for these factors  can lead to further interesting behavior \cite{Delangre06}.
% Indeed, the case where these time-scales are comparable can lead to interesting behavior\cite{Delangre06}, and motivates further investigation. 
In conclusion, we show that the hydrodynamic instability underlying \monami ~differs from the traditional KH due to the presence of the vegetation drag. 
The threshold flow condition observed in the field and in lab experiments arises due to the presence of this drag. 
Furthermore, our linear stability analysis reveals two modes, namely Mode 1 with a mechanism similar to KH and Mode 2 with characteristics distinctly different from KH. 
% While further investigation is needed to understand the sensitivity of the results to the various simplifying assumptions made in our model, the agreement with experiments is encouraging.
The spatial structure of the instability modes has direct implications for transport in the grass bed.
% Mode 1 instability likely leads to enhanced transport near the grass tips as has been observed \cite{Nepf04,Okamoto12}, while Mode 2 instability influences the whole water column.
Our analysis also informs flow structure formation in many other related scenarios, such as flow over coral reefs, permeable sediments, and flow through urban environments, and therefore is expected to have a wider impact.

\acknowledgments
MMB was supported by the Collective Interactions Unit, OIST Graduate University, while visiting Brown University. We thank Heidi Nepf, Marco Ghisalberti, Mitul Luhar, and L. Mahadevan for helpful discussions.

% \bibliography{Grass}{}
% \bibliographystyle{jfm}
\bibliography{Grass}{}

\begin{thebibliography}{10}

\bibitem{AckermanOkubo93}
J.~D. Ackerman and A.~Okubo.
\newblock {Reduced mixing in a marine macrophyte canopy}.
\newblock {\em Functional Ecology}, 7:305--309, 1993.

\bibitem{Chen97}
D.~Chen and G.~H. Jirka.
\newblock {Absolute and convective instabilities of plane turbulent wakes in a
  shallow water layer}.
\newblock {\em Journal of Fluid Mechanics}, 338:157--172, 1997.

\bibitem{Chu91}
V.~H. Chu, {J-H} Wu, and R.~E. Khayat.
\newblock {Stability of transverse shear flows in shallow open channels}.
\newblock {\em Journal of Hydraulic Engineering}, 117(10):1370--1388, 1991.

\bibitem{Drazin81}
P.~G. Drazin and W.~H. Reid.
\newblock {Hydrodynamic stability}.
\newblock {\em Cambridge U. Press, Cambridge}, 1981.

\bibitem{Fonseca87}
M.~S. Fonseca and W.~J. Kenworthy.
\newblock {Effects of current on photosynthesis and distribution of
  seagrasses}.
\newblock {\em Aquatic Botany}, 27(1):59--78, 1987.
\newblock Environmental Impacts On Seagrasses.

\bibitem{Ghisal02}
M.~Ghisalberti and H.~M. Nepf.
\newblock {Mixing layer and cohrent structures in vegetated aquatic flows}.
\newblock {\em J. Geophys. Res.}, 107, 2002.

\bibitem{Nepf04}
M.~Ghisalberti and H.~M. Nepf.
\newblock {The limited growth of vegetated shear layers}.
\newblock {\em Water Resources Research}, 40(7), 2004.

\bibitem{Nepf06}
M.~Ghisalberti and H.~M. Nepf.
\newblock {The sructures of shear layer in flows over rigid and flexible
  canopies}.
\newblock {\em Enviornmental Fluid Mechanics}, pages 277--301, 2006.

\bibitem{Grizzle96}
R.~E. Grizzle, F.~T. Short, C.~R. Newell, H.~Hoven, and L.~Kindblom.
\newblock {Hydrodynamically induced synchronous waving of seagrasses:
  {\lq}monami{\rq} and its possible effects on larval mussel settlement}.
\newblock {\em Journal of Experimental Marine Biology and Ecology},
  206(1--2):165--177, 1996.

\bibitem{Ikeda96}
S.~Ikeda and M.~Kanazawa.
\newblock {Three-dimensional organized vortices above flexible water plants}.
\newblock {\em Journal of Hydraulic Engineering}, 122(11):634--640, 1996.

\bibitem{Inoue56}
E.~Inoue.
\newblock {Studies of the Phenomena of Waving Plants Caused by Wind}.
\newblock {\em J. of Agri. Meteor.}, 11(4):147--151, 1956.

\bibitem{Nepf99}
H.~M. Nepf.
\newblock {Drag, turbulence, and diffusion in flow through emergent
  vegetation}.
\newblock {\em Water resources research}, 35(2):479--489, 1999.

\bibitem{Nepf2012}
H.~M. Nepf.
\newblock {Flow and transport in regions with aquatic vegetation}.
\newblock {\em Annual Review of Fluid Mechanics}, 44:123--142, 2012.

\bibitem{Nepf00}
H.~M. Nepf and E.~R. Vivoni.
\newblock {Flow structure in depth-limited vegetated flow}.
\newblock {\em J. Geophys. Res.}, 105:28,547--28,557, 2000.

\bibitem{Nepf07}
H.~M. Nepf, B.~White, and E.~Murphy.
\newblock {Retention time and dispersion associated with submerged aquatic
  canopies}.
\newblock {\em Water Resources Research}, 43, 2007.

\bibitem{Okamoto12}
T.~Okamoto, I.~Nezu, and H.~Ikeda.
\newblock {Vertical mass and momentum transport in open-channel flows with
  submerged vegetations}.
\newblock {\em Journal of Hydro-environment Research}, 6(4):287--297, 2012.

\bibitem{Delangre06}
C.~Py, E.~{De Langre}, and B.~Moulia.
\newblock {A frequency lock-in mechanism in the interaction between wind and
  crop canopies}.
\newblock {\em J. Fluid Mech.}, 568:425--449, 2006.

\bibitem{Raupach96}
M.~R. Raupach, J.~J. Finnigan, and Y.Brunet.
\newblock {Cohrent eddies and turbulence in vegetation canopies: The mixing
  layer analogy}.
\newblock {\em Boundary-Layer Meteorology}, 78:351--382, 1996.

\bibitem{Rayleigh1879}
{Rayleigh}.
\newblock {On the stability, or instability, of certain fluid motions}.
\newblock {\em Proceedings of the London Mathematical Society}, 1(1):57--72,
  1879.

\bibitem{Vivoni98}
E.~R. Vivoni.
\newblock {Turbulence Structure of a Model Seagrass Meadow}.
\newblock Master's thesis, MIT, 1998.

\bibitem{White07}
B.~L. White and H.~M. Nepf.
\newblock {Shear instability and coherent structures in shallow flow adjacent
  to a porous layer}.
\newblock {\em Journal of Fluid Mechanics}, 593:1--32, 2007.

\end{thebibliography}
\bibliographystyle{plain}
\end{document}